\documentclass[epsfig,11pt]{article}
\usepackage{epsf}
\usepackage{graphicx}
\usepackage{amsmath}
\title{\bf Off-shell gluon production in interaction of a projectile
with 2 or 3 targets}
\author{M.A.Braun and M.Yu.Salykin\\
Dep. of High Energy physics,
 Saint-Petersburg State University,\\
198504 S.Petersburg, Russia}
\def\beq{\begin{equation}}
\def\eeq{\end{equation}}
\def\lra{\leftrightarrow}

\begin{document}
\maketitle
\medskip

\abstract
{Within the effective QCD action for the Regge kinematics amplitudes for virtual gluon
emission are studied in collision of a projectile with two and three targets. It is demonstrated
that all un-Feynman singularities cancel between induced vertices and rescattering contributions.
Formulas simplify considerably in a special gauge which is a straightforward generalization of the
light-cone gauge for emission of real gluons.}

\section{Introduction}
In the framework of the
perturbative QCD at high energies and in Regge kinematics strong interactions
can be described in terms of
reggeized gluons ("reggeons"), which combine into colorless pomerons
exchanged between colliding hadrons. Reggeons and their interactions
were first introduced in the dispersion approach, using multiple unitarity cuts
~\cite{bfkl, bartels}. Later, mostly to describe next order contributions,
a convenient and powerful method of effective action was proposed
in which the reggeons figure as independent dynamical fields interacting with
the standard gluons ~\cite{lipatov}.The effective action allows to present
scattering amplitudes in the Regge kinematics as a sum of diagrams, similar to the
Feynman ones with certain rules for propagators and interaction vertices
~\cite{antonov}. The latter, apart from the standard QCD vertices, include the so-called induced
vertices in which the reggeons interact with two or more gluons. In the effective action approach
the scattering amplitudes depend not only on the transversal variables but also on the
longitudinal ones. So one has to perform the longitudinal integrations to reduce the result to the
purely transverse form, as in the dispersion approach mentioned above. In applications to the
amplitudes with many in-coming or out-going reggeons these integrations are not trivial, since
the induced vertices contain singularities in longitudinal variables different from the standard
Feynman ones. In earlier papers ~\cite{BLSV, BPSV} it was shown however that amplitudes for emission
of a real gluon in transition of a reggeon into two or three reggeons in fact can be rewritten in the
purely transverse form with the standard transversal vertices connected with Feynman propagators.
The contribution from induced vertices becomes substituted by the one from the Feynman propagators of the
rescattering projectile. This result greatly simplifies application of the approach to real processes,
as was illustrated in ~\cite{BPSV1} where scattering off the deuteron projectile was studied.
Note that this result heavily rested on the use of a specific gauge, in which the gluon polarization
vectors were chosen to be orthogonal to the target momentum and in which the relevant vertices radically
simplify.

However in applications (say for the total cross-sections) also vertices for production of a virtual gluon
appear. So the question arises whether the same conclusion holds also in this case. Concretely if also for
the virtual gluon production the contribution from the induced vertices can be traded for the contribution
from the projectile rescattering, thereby liquidating non-Feynman singularities introduced by induced vertices
and reducing longitudinal integration to the standard Feynman ones. This problem is the aim of the present study.
We shall find that the answer is  positive in the sense that the un-Feynman poles are canceled
between the contributions from the vertex and rescattering. There also exists a special gauge in which the bulk of the
result is
reduced to essentially transverse vertices connected by Feynman propagators.
However this also requires certain small change in the transverse vertices and taking a new polarization into account.

These conclusions are rather straightforward in the transition into two out-going reggeons, considered in Sections 2 and 3.
A considerably more complicated case of three out-going reggeons is studied in Sections 4 and 5. Our conclusions
are presented in Section 6.

Note that in ~\cite{kutak1,kutak2} some simple processes initiated by virtual gluons and
essentially mediated by a single reggeon exhchange were studied.
Validity of the effective action technique was confirmed. However reggeon splitting was not
considered, so that the problems treated in our paper were not encountered.

\section{Emission of a virtual gluon in interaction with two targets}
\begin{figure}[h]
\begin{center}
\includegraphics[scale=0.6]{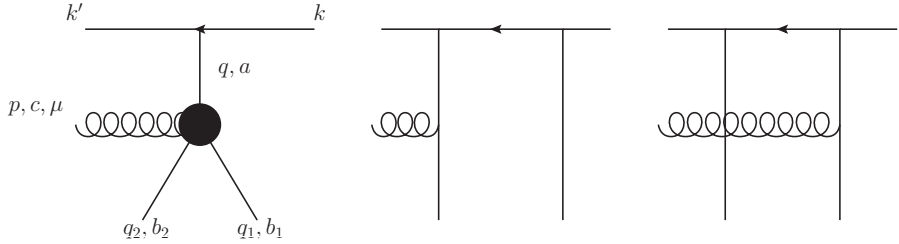}
\end{center}
\caption{Production amplitude: vertex and rescattering contributions}
\label{fig}
\end{figure}

The amplitude  for production of a gluon with momentum $p$, polarization vector $e_\mu$ and
color $c$ in the transition of a reggeon  into two reggeons with momenta $q_1$ and $q_2$
and colors $b_1$ and $b_2$ is represented
by three diagrams shown on Fig. \ref{fig}. The left diagram corresponds
to emission from the effective vertex $V$ for transition of a reggeon to two reggeons plus the virtual gluon
(R$\to$RRP vertex). The other two correspond to
rescattering of the projectile.


\subsection{Emission from the vertex $V$}
The amplitude corresponding to the left diagram in Fig \ref{fig} is given by
\[
{\cal A}^{V}={\cal A}^{V}_\mu e^\mu,\ \
{\cal A}^{V}_\mu=g^3\bar{u}(k')\gamma^+u(k)f^{ab_1d}f^{db_2c}T^a\frac{1}{q_\perp^2}V_\mu .
\]
Here $k$ and $k'$ are the initial and final 4-momenta of the projectile; $q=p+q_1+q_2$, $q_-=0$. We suppress factors
coming from the targets, which have their initial 4-momenta $l$. In the c.m. system both $k$ and $l$
have zero transverse components and $k_-=l_+=0$. $T$ is the quark color matrix.
The R$\to$RRP vertex  is $V=V_1+\Big(1\lra 2\Big)$. In the general gauge and for
arbitrary $p$ part $V_1$ is given by
\beq
V_{1\mu}=\frac{i}{4}\Big\{\frac{a_\mu}{d_1}+\frac{b_\mu}{q_{1-}d_1}+\frac{c_\mu}{q_{1-}}\Big\}
\label{v1},
\eeq
where $d_1$ is the Feynman denominator
$
d_1=(q-q_1)^2+i0
$,
\[ a(p,q_2,q_1)=p_+(3p+4q_1+4q_2)\mu-2p_+^2n_\mu^-\]\beq+
n_\mu^+\Big(-(p+2q_1+q_2,p-q_2)+q_1^2+q_2^2-(p+q_1+q_2)^2-(p+q_2)^2
+2p_+p_-\Big),
\label{a}
\eeq
\beq
b(p,q_2,q_1)=-2(p+q_1+q_2)^2_\perp\Big[(p+2q_2)_\mu-p_+n_\mu^-
+n_\mu^+\Big(\frac{q_2^2}{p_+}+p_-\Big)\Big]
\label{b}
\eeq
and
\beq
c=2n_\mu^-\frac{(p+q_1+q_2)_\perp^2}{p_-}.
\label{c}
\eeq
Here $n^{\pm}=(1,0,0,\mp 1)$.
Note that $Vp=0$ for arbitrary $p$.
Terms in (\ref{v1}) contain poles of the Feynman type at  $(q-q_1)^2=0$ and un-Feynman type at $q_{1-}=0$
coming from the induced vertices.
To separate the latter we use
\beq \frac{1}{q_{1-}d_1}=
\frac{q_+}{(q-q_1)_\perp^2d_1}+\frac{1}{q_{1-}(q-q_1)_\perp^2}.
\label{rest2}
\eeq
Then $V_1$ is transformed into two parts
\beq
V_1=
\frac{i}{4}\Big\{\frac{1}{d_1}\Big( a+\frac{p_+b}{(q-q_1)_\perp^2}\Big)+
\frac{1}{q_{1-}}\Big(\frac{b}{(q-q_1)_\perp^2}+c\Big)\Big\}.
\label{v11}
\eeq
We introduce the generalized  off-mass-shell Lipatov vertex
\beq
2L_\mu(p,q_2)=-\frac{1}{(p+q_2)_\perp^2}\Big\{(p+2q_2)_\mu+\Big(\frac{q_2^2}{p_+}+p_-\Big)n^+_\mu-
\Big(\frac{(p+q_2)_\perp^2}{p_-}+p_+\Big)n^-_\mu\Big\}.
\label{lip}
\eeq
It goes into the standard Lipatov vertex when $p_-+q_{2-}=0$, so that $p_-=-q_{2-}$.
However we preserve this definition also for $p_-+q_{2-}\neq 0$, as in our case when
$p_-+q_{2_-}=-q_{1-}$.
With this definition we find
\beq
\frac{b}{(q-q_1)_\perp^2}+c=4(p+q_1+q_2)^2L(p,q_2).
\label{b1}
\eeq
The first term in (\ref{v11}) may be used as the definition of the  off-mass-shell Bartels vertex
\beq
a+\frac{p_+b}{(q-q_1)_\perp^2}=-4p_+(p+q_1+q_2)^2B(p,q_2,q_1).
\label{defb}
\eeq
so that the vertex $V$ as a whole can be rewritten in the same form as on the mass shell in the gauge $(el)=0$
~\cite{BLSV}
\beq
V_1=i(p+q_1+q_2)^2\Big\{-\frac{p_+B(p,q_2,q_1)}{(p+q_2)^2+i0}+\frac{L(p,q_2)}{q_{1-}}\Big\}.
\label{v12}
\eeq
as well as the amplitude corresponding to emission from the vertex
\beq
{\cal A}^V=ig^3\bar{u}(k')\gamma^+u(k)f^{ab_1d}f^{db_2c}t^a
\Big[-\frac{p_+B(p,q_2,q_1)}{(p+q_2)^2+i0}+\frac{L(p,q_2)}{q_{1-}}\Big] +\Big(1\lra 2\Big).
\label{av}
\eeq
Here the unwanted un-Feynman pole at $q_{1-}=0$ is separated in the second term.

For a real emitted gluon the un-Feynman poles in $V$ were canceled by the contribution from the rescattering diagrams
on the right in Fig. \ref{fig}. Presently we study if this also happens with the virtual emitted gluon.

\subsection{Emission from rescattering}
The two diagrams  in Fig. \ref{fig} on the right describe emission of a gluon during rescattering of the
projectile. If one takes into account the full Feynman projectile propagator they give in the sum
\beq
{\cal A}^{RF}=\frac{g^3\bar{u}(k')\gamma^+u(k)}{q_{2-}+i0}L(p,q_1)f^{b_1cd}t^{b_2}t^{d}+
\frac{g^3\bar{u}(k')\gamma^+u(k)}{-q_{1-}+i0}L(p,q_2)f^{b_2cd}t^{d}t^{b_1}+\Big(1\lra 2\Big).
\label{rescfull}
\eeq
Emission vertex $L(p,q_1)$ is the Lipatov vertex (\ref{lip}) for of-mass-shell gluon.
Note that in contrast to (\ref{lip})  here the sum of "-" components of the two arguments is equal
to zero. However, with our definition in which only $p_-$ is used, these Lipatov vertices coincide with those
in (\ref{av}).

The propagators can be split into the principal value and delta function terms
\[
\frac{1}{\pm q_{1,2-}+i0}={\cal P}\frac{1}{\pm q_{1,2-}}-i\pi\delta(q_{1,2-}).
\]
As was discussed in ~\cite{BLSV}, in the contribution from rescattering one has to keep only the part of the
projectile propagator
containing the $\delta$-function. The part containing the principal value should be dropped.
So the final rescattering contribution is
\beq
{\cal A}^{R}=g^3\bar{u}(k')\gamma^+u(k)\left[-i\pi\delta(q_{2-})L(p,q_1)f^{b_1cd}\Big\{t^{b_2},t^{d}\Big\}-i\pi
\delta(q_{1-})L(p,q_2)f^{b_2cd}\Big\{t^{d},t^{b_1}\Big\}\right].
\label{resc}
\eeq
Expression (\ref{rescfull}) With  Feynman propagators is given by the sum
\beq
{\cal A}^{RF}={\cal A}^R+{\cal A}^{RP},
\label{arrp}
\eeq
where ${\cal A}^{RP}$ contains contributions from the principal value parts of the propagators.
One finds
\beq
{\cal A}^{RP}=ig^3\bar{u}(k')\gamma^+u(k)f^{ab_1d}f^{db_2c}t^a{\cal P}\frac{1}{q_{1-}}L(p,q_1)
+\Big(1\lra 2\Big)
\eeq
It coincides with the second part of (\ref{av}) containing the Lipatov vertices, provided the
singularities at $q_{1.2-}=0$ are taken in the principal value sense.
In the sum of the vertex and rescattering contributions according to (\ref{arrp}) this second part
restores the Feynman propagators in the latter thereby canceling all un-Feynman singularities in the total amplitude,
which becomes
\beq
{\cal A}^{tot}=
-ig^3\bar{u}(k')\gamma^+u(k)f^{ab_1d}f^{db_2c}T^a
\frac{p_+B(p,q_2,q_1)}{(p+q_2)^2+i0}+{\cal A}^{RF}.
\label{atot}
\eeq

So in the end the amplitude does not contain any singularities different from those
provided by the standard Feynman propagators either for the intermediate gluon or for the rescattering quark.

The final expression for the amplitude is rather complicated. However, as for the real emitted gluon,
it drastically simplifies by the choice of a suitable  gauge.

\section{Quasi-lightcone gauge}
Off-mass shell gluons have 3 instead of 2 polarizations.
It is possible to choose  polarization vectors with a minimal
difference as compared to real gluons.
One can choose two of them (transversal) in the same manner as on the mass shell
imposing condition $(e^Tl)=0$ where $l$ is the target momentum. Then
\beq
e^T_+=0,\ \ e^T_-=-\frac{(e^T p)_\perp}{p_+},\ \ \Big(e^T\Big)^2=-1.
\eeq
The third, longitudinal in the 4-dimensional sense, can be chosen as
\beq
e^{L}_\mu=\frac{\sqrt{p^2}}{p_+}n^+_\mu-\frac{1}{\sqrt{p^2}}p_\mu
\eeq
with the properties
\beq
(e^{L}p)=(e^{T}e^L)=0,\ \ \Big(e^L\Big)^2=-1.
\eeq
Vectors $e^T$, $e^L$ together with vector $e^{(0)}_\mu=p_\mu/\sqrt{p^2}$
form a set of 4 independent orthonormalized vectors in the Lopenz space in which any
polarization vector can be expanded. Excluding $e^{(0)}$ for spin 1 particle one can use
$e^T$ and $e^L$ as polarization vectors. We call this gauge quasi-lightcone, having in
mind that the purely transverse polarizations are the same as for the real gluon.

In addition to our previous products with transverse polarizations we have then to
add products with $e^{(L)}$. Due to orthogonality of vertices to $p$
this is equivalent to additional products with $n^+$. Since $(e^Tn^+)=\Big(n^+\Big)^2=0$
all terms in the amplitude proportional to $n^+$ vanish in this gauge. So in the end we can drop all terms
in the vertex containing $p_\mu$ or $n_\mu^+$.
This drastically simplifies the resulting expressions.

Effectively this means that in this gauge we can take
\beq
a_\mu=4p_+(q_1+q_2)_\mu -2p_+^2n_\mu^-,\ \ b_\mu=-2q^2(2q_{2\mu}-p_+n_\mu^-).
\label{abg}
\eeq

First we study contribution from polarization $e^T$.
One finds
\[
(ae^T)=4p_+(qe^T)_\perp,\ \ (be^T)=-4q^2(p+q_2,e^T)_\perp,\ \ (ce^T)=-4q^2\frac{(pe^T)_\perp}{p_+p_-}.
\]
As a result we get from (\ref{defb})
\beq
(Be^T)=\frac{(p+q_2,e^T)_\perp}{(p+q_2)^2_\perp}-\frac{(p+q_1+q_2,e^T)_\perp}{(p+q_1+q_2)^2_\perp}.
\label{be}
\eeq
This is the same expression which one had for the on-mass-shell gluon. So in this quasi-lightcone
gauge, the contribution from the transverse polarizations does not feel the off-mass-shellness of the gluon.
On the other hand we get
\beq
\Big(Le^T\Big)=\frac{(pe^T)_\perp}{p_\perp^2-p^2}-\frac{(q-q_1,e^T)_\perp}{(q-q_1)_\perp^2}.
\label{lip1}
\eeq
Comparing with the real gluon we find a change $p_\perp^2\to p_\perp^2-p^2$ in the denominator.

So for polarizations $e^T$ the changes in the amplitude (\ref{atot}) are minimal: part from the vertex $V$ does not
change at all and in the rescattering part the Lipatov vertex contains $p_\perp^2-p^2$ instead of simply
$p_\perp^2$.

The new parts come from polarization $e^L$. One finds
\[(an^+)=-4p_+^2,\ \ (bn^+)=4q^2p_+,\ \ (cn^+)=4\frac{q^2}{p_-}.\]
This gives
\beq
(Be^L)=\sqrt{p^2}\Big(\frac{1}{q^2}-\frac{1}{(p+q_2)_\perp^2}\Big)
\label{bel}
\eeq
and
for the rescattering
\beq
(Le^L\Big)=\sqrt{p^2}\Big(\frac{1}{(p+q_2)_\perp^2}-\frac{1}{p_\perp^2-p^2}\Big).
\label{lel}
\eeq

This ends the study of the scattering on two centers. We have found that,
first, as for the real emitted gluon all un-Feynman singularities at $q_{1,2}=0$ actually go
when one uses full Feynman propagators for rescattering. Second, in the specially chosen gauge
results for the purely transversal polarizations are nearly identical with the real gluon case
(except for the addition of $-p^2$ to $p_\perp^2$ in the denominators of the Lipatov emission
vertices for rescattering). New contributions from polarization $e^L$ are proportional to $\sqrt{p^2}$
and also contain the additional $-p^2$ in the Lipatov vertices. Apart from this they
do not depend on longitudinal variables.

\section{Interaction with three targets.
The R$\to$RRRP vertex.}
\begin{figure}[h]
\begin{center}
\includegraphics[scale=0.6]{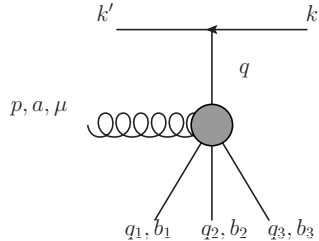}
\end{center}
\caption{The amplitude with the R$\to$RRRP vertex.}
\label{fo}
\end{figure}

\begin{figure}[h]
\begin{center}
\includegraphics[scale=0.6]{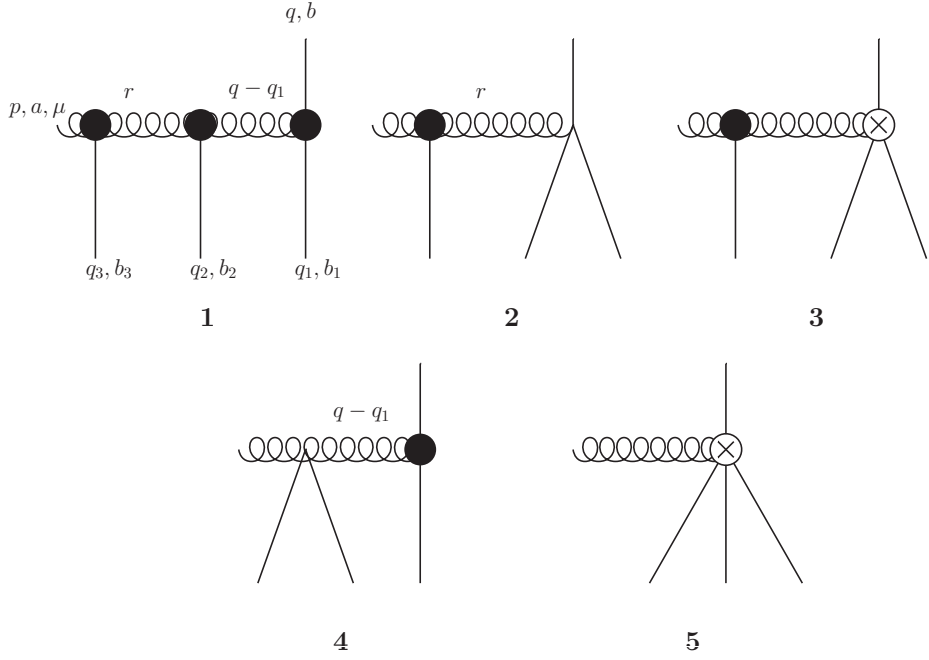}
\end{center}
\caption{Diagrams for the R$\to$RRRP vertex.}
\label{fi}
\end{figure}

In addition to $d_1=(q-q_1)^2+i0$ we introduce $d_2=(q-q_1-q_2)^2+i0$. Here $q=p+q_1+q_2+q_3$.
We shall use (\ref{rest2}) and
and
\begin{equation}
\frac{1}{d_2(q_{1-}+q_{2-})}=
\frac{q_+}{(q-q_1-q_2)^2_\bot d_2}+\frac{1}{(q-q_1-q_2)^2_\bot}\frac{1}{q_{1-}+q_{2-}}
\label{rest2a}
\end{equation}
The amplitude for the gluon emission from the R$\to$RRRP in interaction of the projectile with three targets
is illustrated in Fig. \ref{fo}. The vertex is composed of various contributions
diagrammatically shown in Fig. \ref{fi}. The numbers of diagrams in the following refer to this figure.
Black disks refer to effective vertices in Lipatov's
effective action. Discs with cross indicate the so-called induced vertices. The dot in diagram 2 corresponds
to the QCD 4-gluon coupling.

\subsection{Diagram 1}
We separate the common factor
\beq
F_1=\frac{1}{8}\frac{g^4}{q^2}\bar{u}(k')\gamma_+ u(k)f^{b_3 ad}
f^{b_2 dc}f^{bb_1 c}t^b
\label{fac1}
\eeq
The expression for the diagram can then be written as
\beq
D_1=-\frac{A_1}{d_1d_2}-
\frac{B_1}{q_{1-}d_1d_2}.
\label{d11}
\eeq
Here
\[
A_1=p_+^2(8q-p)_\mu-4p_+^3n_\mu^-
+p_+n_\mu^+\Big[(q_2+q_3-3p)(q+q_1)\]\beq
+q_2^2+2q_3^2+q_1^2+q^2-p^2-
(p-q_3,p+2q_2+q_3)+4p_+p_-\Big]
\label{a1}
\eeq
and
\[
B_1=q^2\Big\{2p_+(-2r-4q_2-p-2q_3)_\mu+4p_+^2n_\mu^-\]\beq
+n_\mu^+\Big[-4p_+4p_-+6p^2-2(p-q_2-q_3)^2-2(q-q_1-q_2)^2\Big]
\Big\}.
\label{b11}
\eeq
Using (\ref{rest2}) we  get
\beq
D_1=
-\frac{1}{d_1d_2}\Big(A_1+\frac{q_+B_1}{(q-q_1)_\perp^2}\Big)-
\frac{B_1}{q_{1-}(q-q_{1})^2_\perp d_2}
\label{d12}
\eeq
Note that both $A_1$ and $B_1$ contain terms proportional
to $(q-q_1-q_2)^2$ which cancel one of the denominators.
\subsection{Diagrams 2 and 3}
For diagrams 2 and 3 we have
\beq
D_{2+3}=\frac{1}{d_2}
\left(
\frac{4p_+q_\bot^2n_\mu^-}{(q_1+q_2)_-q_{1-}}-\frac{4q_\bot^2(p+2q_3)_\mu}{(q_1+q_2)_-q_{1-}}-
\frac{2(p-q_3)_-q_\bot^2n_\mu^+}{(q_1+q_2)_-q_{1-}}-\frac{4q_3^2q_\bot^2n_\mu^+}{p_+(q_1+q_2)_-q_{1-}}+2p_+n_\mu^+
\right).
\label{d23}
\end{equation}
We use $q_{3-}=-(p+q_1+q_2)_-$. Then we find three types of terms:
\beq
D_{2+3}=\frac{A_2}{d_2}+\frac{B_2}{q_{1-}d_2}+
\frac{C_2}{q_{1-}(q_{1-}+q_{2-})d_2},
\label{d231}
\eeq
where
\[
A_2=2p_+n_\mu^+,\ \ B_2=-2q^2n_\mu^+,
\]
\beq
C_2=q^2_\perp\Big\{-4(p+2q_3)_\mu+
n_\mu^+\Big(-4p_--4\frac{q_3^2}{p_+}\Big)
+4p_+n_\mu^-\Big\}.
\label{b_2}
\eeq

Using (\ref{rest2a}) we get
\beq
D_{2+3}=\frac{A_2}{d_2}+\frac{1}{q_{1-}d_2}
\Big(B_2+\frac{q_+C_2}{(q-q_1-q_2)_\perp^2}\Big)+
\frac{C_2}{q_{1-}(q_1+q_2)_-(q-q_1-q_2)_\perp^2}
\label{d232}
\eeq
\subsection{Diagram 4}
Diagram 4 is
\begin{equation}
D_4=\frac{1}
{d_1}
\left(p_+n_\mu^+-\frac{2q_\bot^2n_\mu^+}{q_{1-}}.
\right)
\label{d4}
\end{equation}
It  contains two types of terms
\beq
D_4=\frac{A_4}{d_1}+\frac{B_4}{q_{1-}d_1},
\label{d41}
\eeq
where
\beq
A_4=p_+n_\mu^+,\ \ B_4=-2q^2n_\mu^+.
\label{ab4}
\eeq
Applying (\ref{rest2}) we get
\beq
D_4=\frac{1}{d_1}\Big( A_4+\frac{q_+B_4}{(q-q_1)_\perp^2}\Big)
+\frac{B_4}{q_{1-}(q-q_1)_\perp^2}.
\label{d42}
\eeq

\subsection{Diagram 5}
The diagram 5 is proportional to $1/[q_{1-}(q_{1-}+q_{2-})]$
\begin{equation}
D_5=\frac{B_5}{q_{1-}(q_1+q_2)_-},\ \ B_5=4n_\mu^-\frac{q^2}{p_-}.
\label{d5}
\eeq

\subsection{Total result}
The total amplitude ${\cal A}^V$ coming from the vertex splits into three
parts which we denote similarly to the on-mass shell case
\beq
{\cal A}^V=g^4\bar{u}(k')\gamma_+ u(k)f^{b_3 ad}
f^{b_2 dc}f^{bb_1 c}t^b(W_I+Q_I+R_I)\equiv {\cal A}^{VW}+{\cal A}^{VQ}+{\cal A}^{VR}
\eeq
Here
$W_I$ contains only the Feynman propagators whereas $Q_I$ and $R_I$ also contain
un-Feynmam poles.
The Feynman part is
\beq
W_I=\frac{1}{8q^2d_1d_2}
\Big(-A_1-\frac{q_+B_1}{(q-q_1)_\perp^2}\Big)+
\frac{A_{2}}{8q^2d_2}+\frac{1}{8q^2d_1}
\Big(A_4+\frac{q_+B_4}{(q-q_1)_\perp^2}\Big).
\label{w}
\eeq

The part  $Q_I$ contains terms with a pole at $q_{1-}=0$
\beq
Q_I=\frac{1}{8q^2q_{1-}d_2}
\Big\{-\frac{B_1}{(q-q_1)_\perp^2}+\frac{q_+C_2}{(q-q_1-q_2)_\perp^2}+B_2\Big\}
+\frac{B_4}{8q^2q_{1-}(q-q_1)_\perp^2}.
\eeq
Explicitly
\beq
Q_{I}=\frac{T_1+T_2+T_3}{q_{1-}d_2},
\label{q}
\eeq
where
\beq
T_1=\frac{1}{8(q-q_1)_\perp^2}\Big\{p_+(6p+8q_3+8q_2)_\mu-4p_+^2n_\mu^-
+n_\mu^+\Big(4p_+p_--6p^2+2(p-q_2-q_3)^2\Big)\Big\}
\label{t1}
\eeq
\beq
T_2=\frac{1}{8(q-q_1-q_2)_\perp^2}\Big\{-4p_+(2q_3+p)_\mu +4p_+^2n_\mu^-
-4n_\mu^+(p_+p_-+q_3^2)\Big\}
\label{t2}
\eeq
\beq
T_3=-\frac{1}{4}n_\mu^+
\label{t3}
\eeq

$R_I$ contains terms proportional to $1/[q_{1-}(q_{1-}+q_{2-})]$:
\[R_I=\frac{1}{8q^2q_{1-}(q_{1-}+q_{2-})}\Big(\frac{C_2}{(q-q_1-q_2)_\perp^2}+4n_\mu^-\frac{q^2}{p_-}\Big)
\]
Explicitly
\[
R_{I}=
\frac{1}{q_{1-}(q_{1-}+q_{2-})}\,\frac{1}{8q^2(q-q_1-q_2)^2_\bot}\Big\{
-(p+2q_3)_\mu-n_\mu^+\Big(p_-+\frac{q_3^2}{p_+}\Big)+n_\mu^-\Big(p_++\frac{(p+q_3)^2_\perp}{p_-}\Big)
\Big\}\]\beq
=\frac{1}{q_{1-}(q_{1-}+q_{2-})}L(p,q_3),
\label{rest21}
\end{equation}
where we used our definition (\ref{lip}) of the generalized Lipatov vertex.
To these contribution one also has to add 5 other terms with simultaneous permutations of momenta $q_{1,2,3}$
and color indices $b_{1,2,3}$ of the outgoing reggeons.

To finally sum all different contributions in Section 6
it will be convenient to present the color factor
in the form
$$
t^{b}f^{bb_1 c}f^{cb_2 d}f^{db_3 a}
=\frac{1}{i}[t^{b_1},t^{c}]f^{cb_2 d}f^{db_3 a}
=-[t^{b_{1}},[t^{b_{2}},t^{d}]]f^{db_{3}a}
$$
\begin{equation}
=(t^{b_2}t^{d}t^{b_1}+t^{b_1}t^{d}t^{b_2}
-t^{d}t^{b_2}t^{b_1}-t^{b_1}t^{b_2}t^{d})f^{db_3 a} .
\label{e1c}
\end{equation}
\subsection{Relation to the vertex R$\to$RRP}
On the mass shell and in the gauge $(el)=0$ the vertex R$\to$RRRP
could be related to the vertex R$\to$RRP. In fact we had then
\beq
W_I=\frac{p_+^2B(p,q_2+q_3,q_1)}{[(q-q_1)^2+i0][(q-q_1-q_2)^2+i0]},
\label{wb}
\eeq
\beq
Q_I=-\frac{p_+B(p,q_3,q_2)}{q_{1-}[(q-q_1-q_2)^2+i0]}
\label{qb}
\eeq
and
\beq
R_I=\frac{L(p,q_3)}{q_{1-}(q_{1-}+q_{2-})}
\label {rl}
\eeq
where $B$ and $L$ were the standard Bartels and Lipatov vertices.
The cancelation
of all un-Feynman singularities heavily relied on the relations (\ref{qb})
and (\ref{rl}).

As we have found relation (\ref{rl}) indeed holds also for the off-mass-shell
gluon and in the arbitrary gauge. So our central problem is to  see if (\ref{qb}) is also
valid for the off-shell gluon in the arbitrary gauge.

From our study of the R$\to$RRP vertex we determined (in fact defined) the
generalized Bartels vertex $B$ by Eq. (\ref{defb}). It  follows that
\beq
-p_+B(p,q_3,q_2)=\frac{1}{4(p+q_2+q_3)_\perp^2}\Big(a(p,q_3,q_2)+\frac{p_+b(p,q_2,q_3)}
{2(p+q_3)_\perp^2}\Big).
\eeq
In fact we have seen that
\[p_+b=-2p_+(p+q_2+q_3)_\perp^2\beta,\]
where
\[\beta=(p+2q_3)_\mu-p_+n_\mu^-+n_\mu^+\Big(\frac{q_3^2}{p_+}+p_-\Big).\]
So we get
\beq
-p_+B(p,q_3,q_2)=\frac{a}{4(p+q_2+q_3)_\perp^2}-\frac{p_+\beta}{2(p+q_3)^2}.
\label{pb}
\eeq
It contains two terms with different singularities in the transverse space.
Our expression for $Q_I$ also has two terms with the same  singularities plus
a term with no singularities whatsoever, which can be included in any of the two
previous ones with the appropriate factor.

Forgetting this factor for a while have therefore to compare
$T_1$ with $a/(p+q_2+q_3)_\perp^2$ and $T_2$ with $b/(p+q_3)_\perp^2$
We start from $T_2$. Presenting
\[T_2=\frac{t_2}{8(p+q_3)_\perp^2}\]
we have
\[t_2=-4\Big(p_+(p+2q_3)_\mu -p_+^2n_\mu^-
+n_\mu^+(p_+p_-+q_3^2\Big)\]
to be compared with the second term in (\ref{pb}). We find
that they are identical, so that the part containing $1/(p+q_3)_\perp^2$  in $Q_I$
has indeed the form (\ref{qb}).

Now we compare terms containing $1/(p+p_2+p_3)_\perp^2$
We present
\[T_1+T_3=\frac{t_1}{8(p+q_2+q_3)_\perp^2},\]
where from our previous calculations
\beq
t_1=2p_+(3p+4q_2+4q_3)_\mu-4p_+^2n_\mu^-
+n_\mu^+\Big(4p_+p_--6p^3+2(p-q_2-q_3)^2-2(p+q_2+q_3)_\perp^2\Big).
\eeq
Taking into account factor $1/8$ and $1/4$ in $p_+B$, $t_1$ is to be compared with
\[
2a(p,q_3,q_2)=2p_+(3p+4q_2+4q_3)_\mu-4p_+^2n_\mu^-\]\beq
+2n_\mu^+\Big(-(p+2q_2+q_3,p-q_3)+q_2^2+q_3^2-(p+q_2+q_3)^2_\perp-(p+q_3)^2
+2p_+p_-\Big).
\eeq
We observe that the first two terms are  identical.
The coefficients before $n_\mu^+$ are:
in $t_1$
\[-6p^2+2(p-q_2-q_3)^2-2(p+q_2+q_3)_\perp^2+4p_+p_-\]
and in $2a$
\[
-2(p+2q_2+q_3,p-q_3)+2q_2^2+2q_3^2-(p+q_3)^2
-2(p+q_2+q_3)^2_\perp+4p_+p_-\]
The last two terms are identical. The first ones give
in $t_1$
\[-4p^2+2q_2^2+2q_3^2-4pq_2-4pq_3+4q_2q_3\]
and in $2a$
\[-2p^2-4pq_2-2pq_3+2pq_3+4q_2q_3+2q_3^2+2q_2^2+2q_3^2
-2p^2-2q_3^2-4pq^3\]
These two expressions also coincide.

So we have proven that (\ref{qb}) is also true for the emission
of a virtual gluon in an arbitrary gauge.
This opens the way to demonstrate that all un-Feynman singularities
in ${\cal A}^V$ are canceled by the rescattering contributions,
provided one takes full Feynman quark propagators in them and
treats appropriately the singularities at $q_{i-}=0$ and
$q_{i-}+q_{k-}=0$ with $i\neq k=1,2,3$.

As to relation (\ref{wb})inspection of our results shows that it does not hold generally.
However in the quasi-lightcone gauge it is fulfilled indeed.
In fact in that gauge, as mentioned, we can drop all terms with $p_\mu$ and $n_\mu^+$.
Then one finds
\beq A_{1\mu}=8p_+(q_1+q_2+q_3)_\mu -4p_+^2n_\mu^-,\ \
B_{1\mu}=4q_2\Big(2(q_2+q_3)_\mu-2p_+n_\mu^-\Big).
\label{abg1}
\eeq
Comparison with (\ref{abg}) shows that this corresponds to changing $q_2\to q_2+q_3$ from which
(\ref{wb}) indeed follows in the quasi-lightcone gauge.

\section{Rescattering}
In the rescattering the gluon is emitted by vertices
R$\to$RRP and R$\to$RP (Lipatov) which are known for
off-mass-shell emitted gluon. The diagrams themselves
are quite obvious and do not differ from the on-mass-shell
case. So for calculating the rescattering contribution we can use our results
derived for on-mass-shell emission in ~\cite{BPSV} substituting for
the Bartels and Lipatov vertices their expression for the off-mass-shell
gluon and in the arbitrary gauge introduced in our Section 2. Some details of the
manipulations with color factors can also be found in ~\cite{BPSV}.

\begin{figure}[h]
\begin{center}
\includegraphics[scale=0.60]{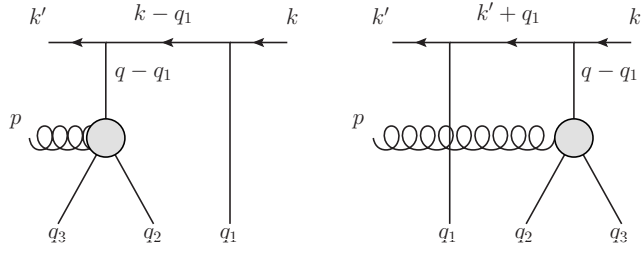}
\end{center}
\caption{Diagrams with the double interaction of the projectile
and the gluon emitted from the R$\to$RRP vertex.}
\label{fii}
\end{figure}

\subsection{Single rescattering. Diagrams in Fig. \ref{fii}}

\begin{figure}[h]
\begin{center}
\includegraphics[scale=0.60]{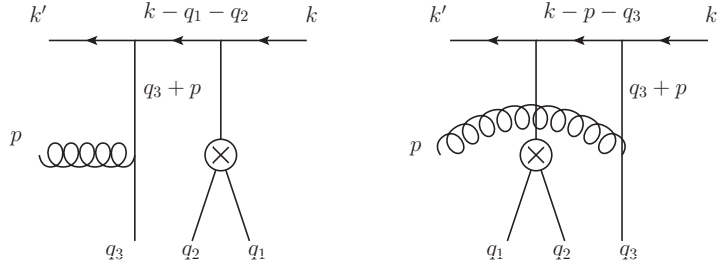}
\end{center}
\caption{Diagrams with the double interactions of the projectile and
the gluon emitted from the R$\to$RP vertex.}
\label{fiv}
\end{figure}

Diagrams with a single rescattering of  the projectile separate into two
groups with emission of the gluon from the R$\to$RRP vertex
shown in Fig. \ref{fii} and from a reggeon shown in Fig. \ref{fiv}.

In both diagrams grey disks correspond to the R$\to$RRP vertex
given analytically in (\ref{v12}).
Since the  vertex is symmetrical with respect to
permutation of the two outgoing reggeons, the symmetrization for
diagrams in Fig.~\ref{fii} has to be carried out only
for cyclic permutations of reggeons 1,2,3.
The color factors are
\begin{equation}
C_1=-i f^{db_3 a} (T^{b_2}T^{d}T^{b_1}-T^{d}T^{b_2}T^{b_1})
\label{e23}
\end{equation}
for the first diagram and
\begin{equation}
C_2=-i f^{db_3 a} (T^{b_1}T^{b_2}T^{d}-T^{b_1}T^{d}T^{b_2})
\label{e24}
\end{equation}
for the second one.

The projectile factors, apart from the color factors, contain the
quark  propagators, which in the Regge limit are
\beq
 \frac{1}{\pm q_{1-}+i0}
 \label{prop}
 \eeq
with the sign "-" for the left diagram and "+" for the right one.
As mentioned in the rescattering only the delta-function part of the
propagator should be left. So  (\ref{prop}) should be substituted as
\beq
 \frac{1}{\pm q_{1-}+i0}\to-i\pi\delta(q_{1-}).
 \label{prop1}
 \eeq

With these preliminaries we find the contribution to the rescattering amplitude from
Fig. \ref{fii} as
\beq
{\cal A}^R_1=g^4\bar{u}(k')\gamma_+u(k)(C_1+C_2)(-i\pi)\delta(q_{1-})
\Big[-\frac{q_+B(p,q_3,q_2)}{(q-q_1-q_2)^2+i0}
+\frac{L(p,q_3)}{q_{2-}}\Big].
\label{ar1}
\eeq

\subsection{Single rescattering. Diagrams in Fig. (\ref{fiv})}

Passing to the diagrams in Fig. \ref{fiv}
we have their color factors
\[C_3=-if^{db_3a}(T^{d}T^{b_1}t^{b_2}-T^{d}T^{b_2}T^{b_1})\]
for the first diagram and
\[C_4=-if^{db_3a}(T^{b_1}t^{b_2}T^{d}-T^{b_2}T^{b_1}T^{d})\]
for the second

The three-reggeon vertex R$\to$RR was calculated in \cite{BLSV},
to be
\begin{equation}
\frac{gf^{cb_1 b_2}}{2q_{1-}} (q_1 + q_2)_{\perp}^{2}
=-\frac{gf^{cb_1 b_2}}{2q_{2-}} (q_1 + q_2)_{\perp}^{2}
\label{e46}
\end{equation}
(we have $q_{1-}+q_{2-}=0$ in the vertex).
The quark propagators are
\[\frac{1}{\pm (q_{1-}+q_{2-})+i0}\]
with signs "-" for the first diagram and "+" for the second.
Taking into account that we have to leave only delta-functional
parts we finally get the rescattering amplitude for Fig. \ref{fiv}
as
\beq
{\cal A}^R_2=g^4\bar{u}(k')\gamma_+u(k)(C_3+C_4)(-i\pi)\delta(q_{1-}+q_{2-})\frac{L(p,q_3)}{q_{1-}}
\label{ar2}+\Big({\rm permutations}\ (123)\Big)
\eeq

\subsection{Double rescattering of  the projectile}

\begin{figure}[h]
\begin{center}
\includegraphics[scale=0.55]{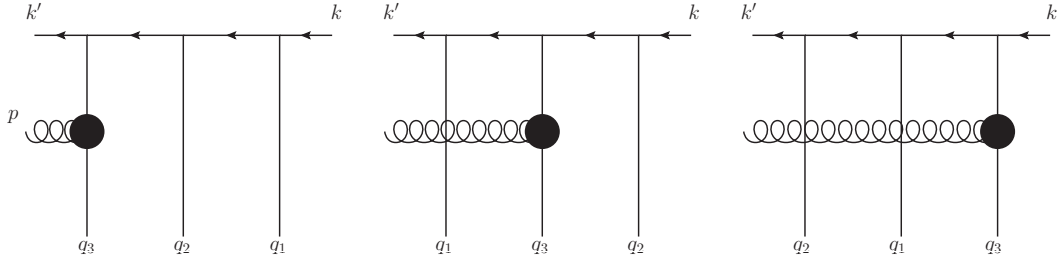}
\end{center}
\caption{Diagrams with the triple interaction of the projectile.}
\label{fiii}
\end{figure}

There are 18 diagrams with three reggeons interacting with the projectile
quark. Three of them are shown in Fig.~\ref{fiii}, the others can be
obtained by means of all permutations of reggeons 1,2,3.
The color factors for the three diagrams in Fig. \ref{fiii}
are
\[C_5=f^{db_3a}T^dT^{b_2}T^{b_1},\ \ C_6=f^{db_3a}T^{b_1}T^dT^{b_2},\ \
C_7=f^{db_3a}T^{b_2}T^{b_1}T^{d}\]
The quark propagators are
\[\frac{1}{-(q_{1-}+q_{2-})+i0}\, \frac{1}{-q_{1-}+i0}\]
in the first diagram
\[\frac{1}{q_{1-}+i0}\,\frac{1}{-q_{2-}+i0}\]
in the second diagram
and
\[\frac{1}{q_{1-}+i0}\,\frac{1}{q_{1-}+q_{2-}+i0}\]
in the third diagram. As it was discussed in ~\cite{BLSV} due to locality in quark rapidity v.p. poles of quark propagators should be dropped.
So in all propagators we are to retain only the $\delta$-functional parts.
Taking into account that
\[\delta(q_{1-})\delta(q_{1-}+q_{2-})=\delta(q_{2-})\delta(q_{1}+q_{2-})=\delta(q_{1-})\delta(q_{2-})\]
we find
the rescattering amplitude
\beq
{\cal A}^R_3=-g^4\bar{u}(k')\gamma_+u(k)L(p,q_3)(C_5+C_6+C_7)(-i\pi)^2\delta(q_{1-}\delta(q_{2-}).
\label{ar3}
\eeq

\section{Cancelation of  un-Feynman singularities}
Inspection of our expressions for the contributions from the R$\to$RRP vertex and
from rescattering we see that they have exactly the same form as in the case
of emitted on-mass-shell gluons in the gauge $(el)=0$. All the difference
is due to the newly defined generalized Lipatov and Bartels vertices. The crucial point is that
the same Bartels vertex $B(p,q_3,q_2)$ appears in the vertex amplitude ${\cal A}^{VQ}$ and
rescattering amplitude ${\cal A}^R_1$, exactly as happens in the on-mass-shell case.
As a result cancelation of un-Feynman singularities and restoration of Feynman propagators for
the rescattering projectile proceeds in the same manner as in the on-mass-shell case
~\cite{BPSV}. So we only briefly comment on the derivation.

Terms containing un-Feynman singularities separate into two groups
depending on their momentum structure. One part contains the
generalized Bartels vertex $B$. Contribution to this group
come from the amplitude ${\cal A}^{VQ}$, Eq. (\ref{q}), and the first term
in the rescattering amplitude ${\cal A}^{R}_1$, Eq.(\ref{ar1}).
Taking into account (\ref{e1c}) we find that
these terms differ only in that factor $\pm 1/q_{1-}$ in ${\cal A}^{VQ}$
is substituted by $-i\pi\delta(q_{1-})$ in ${\cal A}^{R}_1$.
In the sum of these two contributions we get the normal Feynman propagators
\[\pm\frac{1}{q_{1-}}-i\pi\delta(q_{1-})=\frac{1}{\pm q_{1-}+i0}.\]
As a result the contribution from $B(p,q_3,q_2)$ to the total amplitude
will be given by
\beq
{\cal A}^{VQF}_1
g^4\bar{u}(k')\gamma_+ u(k)f^{b_3 cd}\frac{q_+B(p,q_3,q_2)}{(q-q_1-q_2)^2+i0}
\Big[
\frac{if^{ab_2d}T^aT^{b_1}}{-q_{1-}+i0}+\frac{if^{ab_2d}T^{b_1}T^a}{q_{1-}+i0}\Big],
\label{vqf}
\eeq
which is actually the part of the rescattering contribution ${\cal A}^R_1$ with the Bartels vertex and
Feynman propagator for the rescattering projectile.

Thus we are left with only the terms containing the Lipatov vertex $L$.
They come from all contributions. We concentrate on the terms containing
$L(p,q_3)$. We separate the common factor
\[ g^4\bar{u}(k')\gamma_+u(k)f^{db_3 a}L(p,q_3).\]
It has to be multiplied by the sum of different contributions from our
amplitudes. Denoting different terms as $F^I$, $I=V,R1,R2,R3$ for contributions from
amplitudes ${\cal A}^{VR}$, ${\cal A}^{R}_1$, ${\cal A}^{R}_2$ and ${\cal A}^{R}_3$
respectively we find
$$
F^{VR}=
\frac{T^{b_2}T^{d}T^{b_1}+T^{b_1}T^{d}T^{b_2}
-T^{d}T^{b_2}T^{b_1}-T^{b_1}T^{b_2}T^{d}} {q_{1-}(q_{1-}+q_{2-})}
+ \Big(1\lra 2\Big) ,
$$
$$
F^{R1}=
-\frac{T^{b_2}T^{d}T^{b_1}-T^{d}T^{b_2}T^{b_1}
+T^{b_1}T^{b_2}T^{d}-T^{b_1}T^{d}T^{b_2}} {q_{2-}}
(-i\pi)\delta(q_{1-})
+ \Big(1\lra 2\Big) ,
$$
$$
F^{R2}=
\left( \frac{T^{d}T^{b_2}T^{b_1}}{q_{1-}}
-\frac{T^{b_2}T^{b_1}T^{d}}{q_{2-}} \right)
(-i\pi)\delta(q_{1-}+q_{2-})
+ \Big(1\lra 2\Big),
$$
\beq
F^{R3}=
-(T^{d}T^{b_2}T^{b_1}+T^{b_1}T^{d}T^{b_2}+T^{b_2}T^{b_1}T^{d})
(-i\pi)^2 \delta(q_{1-})\delta(q_{2-})
+ \Big(1\lra 2\Big) .
\label{e53}
\end{equation}
Straightforward algebraic manipulations (see ~\cite{BPSV}) demonstrate that the sum of these contributions
leads to the amplitude which corresponds to the double rescattering Fig. \ref{fii} with Feynman propagators
for the rescattering projectile, namely
$$
{\cal A}^{RF}_3=
-g^{4}\gamma_{+}f^{db_3 a}
\left[ \frac{k_{+}^{2}\cdot T^{d}T^{b_2}T^{b_1}}
 { ((k-q_{1}-q_{2})^{2}+i0) ((k-q_{1})^{2}+i0) }
+ \frac{k_{+}^{2}\cdot T^{b_1}T^{d}T^{b_2}}
 { ((k'+q_{1})^{2}+i0) ((k-q_{2})^{2}+i0)}
\right.
$$
\begin{equation}
+ \left. \frac{k_{+}^{2}\cdot T^{b_{2}}T^{b_{1}}T^{d}}
 { ((k'+q_{2})^{2}+i0) ((k'+q_{1}+q_{2})^{2}+i0) } \right]
L(p,q_3)
+ \Big(\mbox{permutations of } 1,2,3\Big) .
\label{e57}
\end{equation}

In conclusion the total amplitude for production of a virtual gluon on three centers
is given by the sum of three contributions
\beq
{\cal A}^{tot}={\cal A}^{VW}+{\cal A}^{VQF} +{\cal A}^{RF}_3
\label{atot3}
\eeq
corresponding to emission from the vertex and single and double rescattering,
in which all propagators are of the Feynman type.

In the general gauge this expression is quite complicated. However it drastically
simplifies in the quasi-lightcone gauge, introduced in Section 3. For transverse polarizations
the Bartels vertex takes its standard form and the Lipatov one only slightly changes.
For longitudinal polarization both vertices become quite simple (Eqs. (\ref{bel}) and (\ref{lel})).

\subsection{Conclusion}
We have found that with the  gluon produced off mass shell both for two and three targets
we can drop the induced part of the effective vertex and use instead
full quark propagators in the rescattering contribution provided we
interpret the singularities at $q_{1-}=0$ and $q_{2-}=0$ in the
induced part in the principal value sense.

However in contrast to the on-mass-sell case
the production amplitude cannot be deduced  from the purely transverse
BFKL-Bartels picture by just introducing Feynman propagators into
the transverse Lipatov and Bartels vertices. Instead one has to
appropriately change these vertices to account for the virtuality of
the emitted gluon, so that they result to  be not purely transversal.

A remarkably  simple result follows in a particular gauge, which is an immediate
generalization of the light-cone gauge for the off-mass-shell gluon. Then the
changes in the vertices is minimal. Still a new, longitudinal,
polarization is to be taken to account.

\section{Acknowledgements}
The authors acknowledge Saint-Petersburg State University for a research grant 11.38.223.2015 and RFFI for a research grant 15-02-02097A.

\end{document}